\documentclass[aps,prb,twocolumn,reprint,superscriptaddress,letterpaper
]{revtex4-2}
\usepackage{filecontents}
\usepackage{graphicx}
\usepackage{bm}
\usepackage{amssymb}
\usepackage{color}
\usepackage{amsmath}
\usepackage{ulem}
\usepackage{dcolumn}
\usepackage{mathrsfs}
\usepackage{url}
\usepackage[
bookmarks=true,
colorlinks,
linkcolor=blue,
urlcolor=blue,
citecolor=blue,
plainpages=false,
pdfpagelabels,
final,
breaklinks=true
]{hyperref}

\begin{document}

\preprint{APS/123-QED}

\title{Higher-order Dirac Semimetal in a Photonic Crystal}

\author{Zihao Wang}
\email{These authors contributed equally to this work}
\affiliation{Division of Physics and Applied Physics, School of Physical and Mathematical Sciences, Nanyang Technological University, 21 Nanyang Link, Singapore 637371, Singapore}

\author{Dongjue Liu}
\email{These authors contributed equally to this work}
\affiliation{School of Electrical and Electronic Engineering, Nanyang Technological University, 50 Nanyang Avenue, Singapore 639798, Singapore}

\author{Hau Tian Teo}
\affiliation{Division of Physics and Applied Physics, School of Physical and Mathematical Sciences, Nanyang Technological University, 21 Nanyang Link, Singapore 637371, Singapore}

\author{Qiang Wang}
\affiliation{Division of Physics and Applied Physics, School of Physical and Mathematical Sciences, Nanyang Technological University, 21 Nanyang Link, Singapore 637371, Singapore}

\author{Haoran Xue}
\email{haoran001@e.ntu.edu.sg}
\affiliation{Division of Physics and Applied Physics, School of Physical and Mathematical Sciences, Nanyang Technological University, 21 Nanyang Link, Singapore 637371, Singapore}

\author{Baile Zhang}
\email{blzhang@ntu.edu.sg}
\affiliation{Division of Physics and Applied Physics, School of Physical and Mathematical Sciences, Nanyang Technological University, 21 Nanyang Link, Singapore 637371, Singapore}
\affiliation{Centre for Disruptive Photonic Technologies, The Photonics Institute, Nanyang Technological University, 50 Nanyang Avenue,
Singapore 639798, Singapore}

\date{\today}

\begin{abstract}
The recent discovery of higher-order topology has largely enriched the classification of topological materials. Theoretical and experimental studies have unveiled various higher-order topological insulators that exhibit topologically protected corner or hinge states. More recently, higher-order topology has been introduced to topological semimetals. Thus far, realistic models and experimental verifications on higher-order topological semimetals are still very limited. Here, we design and demonstrate a three-dimensional photonic crystal that realizes a higher-order Dirac semimetal phase. Numerical results on the band structure show that the designed three-dimensional photonic crystal is able to host two four-fold Dirac points, the momentum-space projections of which at an edge are connected by higher-order hinge states. The higher-order topology can be characterised with the calculation of the $\chi^{(6)}$ topological invariant at different values of $k_z$. An experiment at microwave frequencies is also presented to measure the hinge state dispersion. Our work demonstrates the physical realization of a higher-order Dirac semimetal phase and paves the way to exploring higher-order topological semimetals phases in three-dimensional photonic systems.
\end{abstract}

\maketitle

Band topology has been a major topic in condensed matter physics since the discovery of the quantum Hall effect \cite{klitzing1980new}, leading to discoveries of various topological phases of matter \cite{hasan2010colloquium,qi2011topological}. In recent years, the notion of band topology has also been utilised to manipulate classical waves, such as in the emerging fields of topological photonics \cite{lu2014topological,khanikaev2017two,ozawa2019topological} and topological acoustics \cite{yang2015topological,ma2019topological}. In both electronic and classical topological systems, the bulk-boundary correspondence dictates the connection between the bulk topology and the topological boundary states. Conventionally, the dimension of topological boundary states is one order lower than that of the bulk. For example, a two-dimensional (2D) Chern insulator supports chiral edge states that propagate along its one-dimensional (1D) edges.

Recently, it has been found that the dimension difference between the bulk and the topological boundary states can be larger than one, leading to the discovery of higher-order topological phases \cite{benalcazar2017quantized,benalcazar2017electric,langbehn2017reflection,song2017d,schindler2018higher,ezawa2018higher}. For example, a 2D second-order topological insulator (TI) does not host protected 1D edge states. Instead, it supports zero-dimensional (0D) topological corner states localized at corners. Over the last four years, various higher-order topological systems have been theoretically proposed and experimentally demonstrated \cite{langbehn2017reflection,song2017d,schindler2018higher,imhof2018topolectrical,serra2018observation,peterson2018quantized,ezawa2018higher,noh2018topological,xue2019acoustic,ni2019observation,zhang2019second}. So far, most studies have focused on higher-order TIs, i.e., systems with a bulk bandgap.

Topological semimetals (TSMs) are another class of topological phases \cite{armitage2018weyl}. Different from TIs, TSMs host gapless bulk band structures. One paradigmatic example is the Weyl semimetal, which supports two-fold linear band crossings called Weyl points in the bulk, and topological Fermi-arc states on the surfaces connecting two Weyl points with opposite topological charges \cite{wan2011topological}. Very recently, studies have found that TSMs can also support higher-order topological states. Theoretical proposals have predicted topological hinge states in higher-order Weyl semimetals, Dirac semimetals and nodal lines semimetals \cite{wan2011topological,liu2014discovery,lu2015experimental,xu2015discovery,bernevig2015s,lu2015experimental,burkov2016topological,fang2016topological,lin2018topological,armitage2018weyl,wieder2020strong,luo2021observation,wei2021higher,qiu2020higherorder}. A few experiments have been performed on the platform of acoustic metamaterials to characterize higher-order TSM phases \cite{luo2021observation,wei2021higher,qiu2020higherorder}.

In this work, we introduce the concept of higher-order TSMs into photonic systems by presenting the design and realization of a higher-order Dirac semimetal (HODSM) in a three-dimensional (3D) photonic crystal. Our model consists of coupled layers of deformed photonic honeycomb lattices. We identify the Dirac points (DPs), topology of bulk bands, and the existence of hinge states through full-wave simulations and tight-binding calculations. In addition, a microwave experiment on the dispersion of the hinge states is also presented.

\begin{figure}
\includegraphics{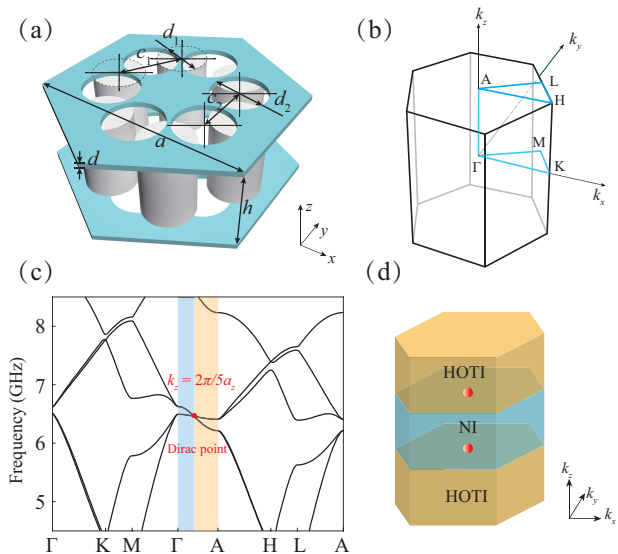}
\caption{\label{tu1} Structure and bulk dispersion of the photonic crystal exhibiting a higher-order Dirac semimetal phase. (a) Schematic of the unit cell of a deformed honeycomb lattice with lattice constant $a$. The lattice consists of grey rods sandwiched by metallic plates. The rods have height $h$, diameter $d_1$ and are separated by a distance $c_1$. The two metallic plates on the top and bottom are drilled with holes to couple the neighbouring layers. The holes have the depth $d$, diameter $d_2$ and are separated by a distance $c_2$. (b) The first Brillouin zone of the crystal. (c) Simulated bulk dispersion of the photonic crystal. The coloured region from $\Gamma$ to A indicates the topological transition when $k_z$ goes through the Dirac point at $k_z$ = $2\pi/5a_z$. (d) Schematic of phase transition in the entire Brillouin zone at different values of $k_z$. The upper and lower parts host the higher-order topological insulator (HOTI) phase, while the middle part possesses the normal insulator (NI) phase. The two red dots denote the two Dirac points.}
\end{figure}

The designing principles of our photonic HODSM are as follows. First, let us consider a 2D photonic honeycomb lattice with each lattice site occupied by a dielectric rod and the surrounding space being air. Then, we choose a larger cell that contains six rods (see Fig.~1(a)) and moves the rods towards/away from the centre of the cell \cite{wu2015scheme}. Previous studies have shown that both deformations will open a bandgap at $\Gamma$ point. When the configuration changes from one deformation to the other, the bandgap closes and reopens, inducing the phase transition between a normal insulator (NI) phase and a higher-order topological insulator (HOTI) phase \cite{xie2018second,zangeneh2019nonlinear,xie2019visualization,liu2019helical,benalcazar2019quantization,li2020higher,xie2020higher}. As we adjust the distance between rods, the coupling strength between them changes as well. For the original honeycomb lattice, the distance of two adjacent rods within a unit cell and that between two unit cells are equal, thus leading to the equal intracell and intercell coupling. In our photonic lattice, we adopt the shrunken lattice in which the distance of rods in a unit cell is smaller, so the ratio between the intercell and intracell coupling is tuned to be less than one. As a result, a complete bandgap is opened at $\Gamma$ point in the Brillouin zone (BZ) and the 2D lattice is in the NI phase. The final step is to extend this 2D lattice into a 3D one by properly engineering the interlayer couplings. The key of our construction is, as can be seen from Fig.~1(a), a rod not only couples to rods at the same in-plane position in adjacent layers but also couples to rods at neighbouring in-plane positions within the same unit cell in adjacent layers. Thus, a $k_z$-dependent term is introduced as a tuning parameter in the Bloch Hamiltonian as an additional term, changing the intercell and intracell coupling strengths as a function of $k_z$. In this model, the situation with equal intercell and intracell coupling strengths is a transition point and it corresponds to a DP in the momentum space. The NI phase and the HOTI phase are separated by the DP, forming a TSM. The tight-binding model analysis is performed in the Supplementary Material \uppercase\expandafter{\romannumeral1} \cite{SI} to show the details.

Fig.~\ref{tu1}(a) shows the designed unit cell, which consists of six dielectric cylindrical rods. These rods are made of dielectric material that has the relative permittivity $\epsilon_r$ = 8.5 and relative permeability $\mu_r$ = 1. Two metallic plates drilled with air holes are placed on the top and at the bottom, acting as the perfect electric conductor (PEC) and the air holes enable wave coupling between adjacent layers. The leftover part of the unit cell is filled with air with $\epsilon_r$ = 1 and $\mu_r$ = 1. The in-plane lattice constant is $a$ = 25 mm. In the vertical direction, the lattice constant is $a_z$ = 10.2 mm and the height of the rods is $h$ = 9 mm. The metallic plates have thickness $d$ = 0.6 mm and thus the air holes between layers are 1.2 mm thick. The rods have the diametre $d_1$ = 6 mm and the air holes $d_2$ = 6.4 mm. The distance between the adjacent rods centres is $c_1$ = 7.88 mm, and the distance between adjacent air holes is $c_2$ = 7.94 mm. Numerically obtained bulk dispersion is shown in Fig.~\ref{tu1}(c). The reciprocal lattice is depicted in Fig.~\ref{tu1}(b) with the first BZ identified using blue lines. In Fig.~\ref{tu1}(c), the band dispersion of the left region ($\Gamma$-K-M-$\Gamma$) lies on the $k_z$ = 0 plane, holding a topologically trivial phase, while the right region (A-H-L-A) lies on the $k_z$ = $\pi/a_z$ plane, holding a topologically nontrivial phase. The middle region ($\Gamma$-A) shows the transition of $k_z$ from $k_z$ = 0 to $\pi/a_z$. The DP appears at $k_z$ = $2\pi/5a_z$, serving as the phase transition point dividing the left and right halves of the bulk dispersion diagram. Fig.~\ref{tu1}(d) depicts the distribution of the DPs along $k_z$ in the BZ. The upper and lower parts of the BZ hold the topologically nontrivial HOTI phase, while the middle part between two DPs has the topologically trivial NI phase. Both in Fig.~\ref{tu1}(c) and (d), the same colours are used to identify the same phases along the phase transition process.


\begin{figure*}
\includegraphics{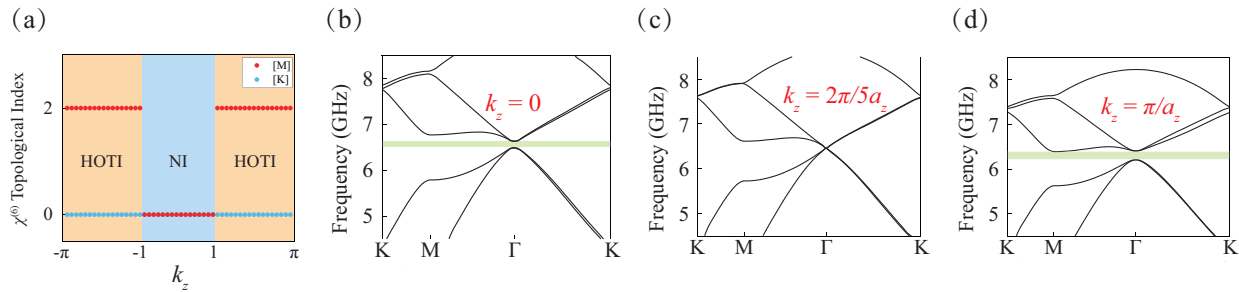}
\caption{\label{tu2} The $\chi^{(6)}$ topological invariants and the band structures for the model in Fig.~\ref{tu1}(a) with different $k_z$ values. (a) The $\chi^{(6)}$ = $([\bold{M}],[\bold{K}])$ topological invariants calculated from the tight-binding model in the first Brillouin zone showing the phase transition. (b-d) The bulk band structure for that in Fig.~\ref{tu1}(a) with $k_z=$0, $2\pi/5a_z$, and $\pi/a_z$ separately. The green bands in (b) and (d) show the bandgaps when $k_z$ deviates from the Dirac point. }
\end{figure*}

To verify the topological nature of the sample, when $k_z$ roams over the first BZ, the $\chi^{(6)}$ topological invariants are calculated to show the phase transition process and the HOTI phase between two DPs. The calculation is done on a series of 2D planes with different $k_z$. As proposed in \cite{noh2018topological}, the phase transition can be represented by checking a pair of topological invariants $\chi^{(6)}$ = $([\bold{M}],[\bold{K}])$. Since the system maintains rotational invariance at several high-symmetry momentum points, it allows us to use the difference of rotation eigenvalues between three momentum points, M, K and $\Gamma$, to characterise the topological nature of that plane. The $[\bold{M}]$ and $[\bold{K}]$ are defined as $[\bold{M}] = \#\mathrm{M_1}-\#\Gamma_1^{(2)}$, $[\bold{K}]=\#\mathrm{K_1}-\#\Gamma_1^{(3)}$, and $[\bold{M}], [\bold{K}]\in \mathbb{Z}$. Here, $\#\mathrm{M_1}$ is the number of bands below the gap in the spectrum which have the rotation eigenvalue 1 at this rotation invariant momenta point M and similar for $\#\Gamma_1^{(\alpha)}$, $\alpha=2,3$, and $\#\mathrm{K_1}$ (See Supplementary Material \uppercase\expandafter{\romannumeral2} \cite{SI} for details). As shown in Fig.~\ref{tu2}(a), when $k_z$ varies, $[\bold{K}]$ will stick to 0 all the time, but $[\bold{M}]$ experiences two times of jumps, which exactly corresponds to the critical point when the intracell and intercell coupling strengths are the same. In the HOTI phase, $[\bold{M}]$ holds the value 2 while in the NI phase it jumps to 0. Figs.~\ref{tu2}(b-d) show the band structure of the model in Fig.~\ref{tu1}(a) when $k_z$ equals to 0, $2\pi/5a_z$, and $\pi/a_z$, respectively. The bulk topology experiences a transition here along with the bandgap at $\Gamma$ point closes and reopens, transforming from NI to HOTI as discussed before.

\begin{figure*}
\includegraphics{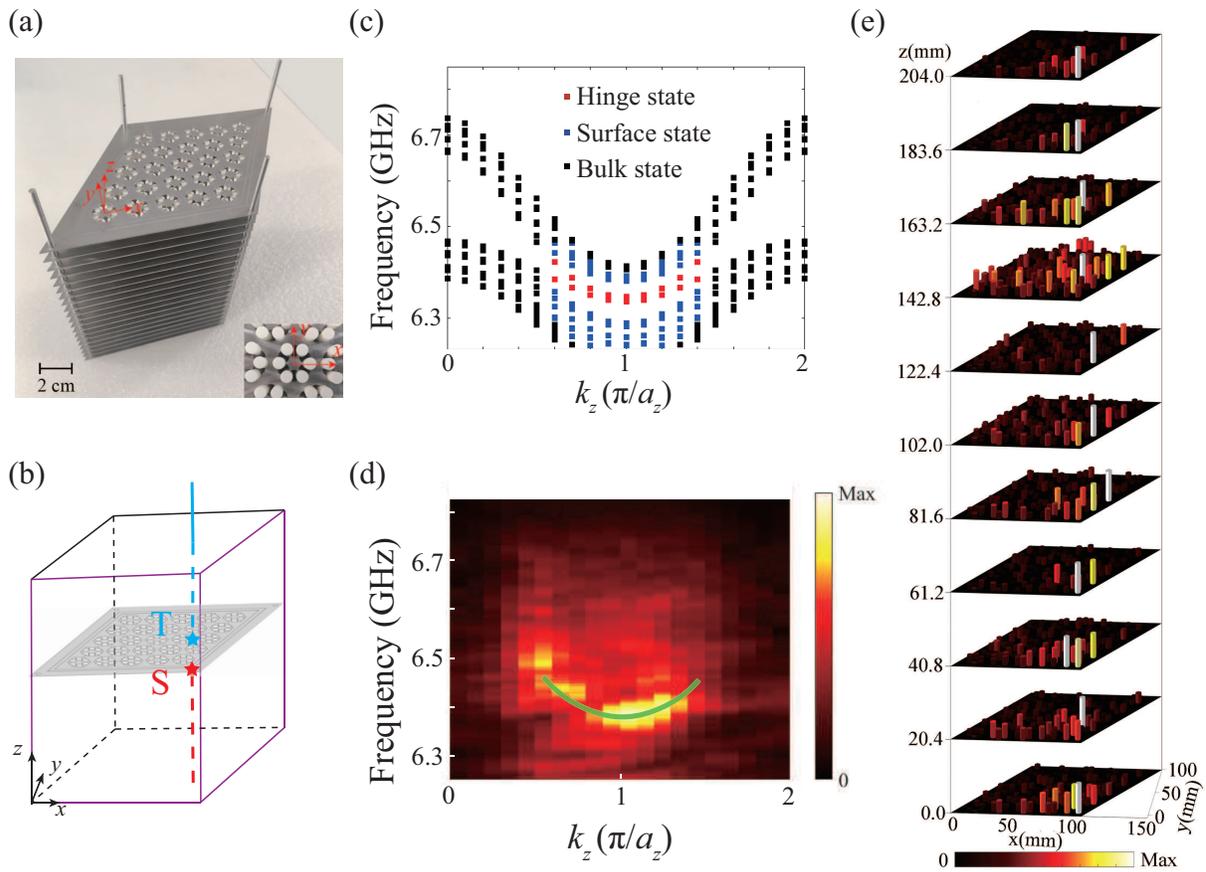}
\caption{\label{tu3} Experimental measurement on the higher-order hinge states. (a) Photograph of the experimental sample. The inset shows the details of the deformed lattice rods placed on each layer with the top plate removed. (b) The sketch of the experimental setup. The red line denotes the source pin inserted from the top air hole. The red star denotes the source dipole fixed at the half-height of the sample next to hard boundaries. The blue line denotes the position of the probe. The purple lines denote the side surfaces with perfect electric conductors and the measurement is performed around the obtuse angle consisting of these two surfaces. (c) Simulation result of a 20 $\times$ 20 lattice. The black, blue, and red dots denote the bulk, surface, and hinge state. (d) The measured energy dispersion. The green curve represents the hinge states redrawn from (c). (e) The measured field distribution showing the hinge state localised on the hard boundary of the sample.}
\end{figure*}

To demonstrate the HODSM and observe the hinge states, we fabricate a 3D sample as shown in Fig.~\ref{tu3}(a). The sample is a stack of 19 layers in total. The four long thin rods at the corners are for positioning and alignment. The inset is the zoom-in view of deformed hexagonal clusters. In the experiment, in order to observe the hinge state localisation, we have adopted PECs to cover the two side surfaces around the obtuse-angle corner to form the hard boundary. (see Fig.~\ref{tu3}(b) and Supplemental Material \uppercase\expandafter{\romannumeral3} \cite{SI}). Fig.~\ref{tu3}(b) illustrates the experimental setup. When probing the hinge state, a source (shown as the red star) is fixed near the hinge located at the hard boundary at the half-height of the whole structure. Then we use a probe (shown as the blue star) to scan over all the unit cells. After collecting the data, the Fourier transform is performed and the result is shown in Fig.~\ref{tu3}(d).

Fig.~\ref{tu3}(c) is the simulation result for a structure with 20 $\times$ 20 unit cells in the $xy$ plane and periodic boundary condition along $z$ direction. The black, blue and red dots represent the bulk, surface and hinge states, respectively. Before the phase transition point at $k_z$ = $2\pi/5a_z$, all the states belong to the bulk states, indicating a NI phase. After the phase transition point, the higher-order hinge states emerge along with the surface states and the hinge states, as predicted by the $\chi^{(6)}$ index. Fig.~\ref{tu3}(d) shows the experimental results of the hinge states. We also redraw the numerical hinge dispersion by the green curve. As can be seen, the experiment result coincides with the simulation result well. Fig.~\ref{tu3}(e) is the field distribution of the hinge state measured from the experiment. Here we show the result of eleven sample planes with an identical height difference of 20.4 mm between any two adjacent planes. The frequency is chosen to be 6.4 GHz, falling in the range of the hinge states dispersion in Fig.~\ref{tu3}(c) and (d). In Fig.~\ref{tu3}(e), the colour and the height of the tiny columns on each layer denote the amplitude of the field at that site point. At each layer, the amplitude of the strongest site is taken as the basis for the normalisation and all the other sites are normalised according to it. As we can find in Fig.~\ref{tu3}(e), all the eleven layers show the trend that the field localises on the hinge on the hard boundary. Compared to the hinge localisation, the states in the bulk away from the corner decreases rapidly and become invisible. The experimental field distribution of the bulk distribution can be found in Supplementary Material \uppercase\expandafter{\romannumeral4} \cite{SI} and obvious contrast between the hinge and bulk mode are shown.

In our sample, it is a parallelogram in the $x$-$y$ plane, so it hosts two obtuse-angle corners and two acute-angle ones. Both the obtuse-angle corners are able to hold the hinge state localisation there. The hinge mode localised on the corner with hard boundaries is discussed above. On the other obtuse angle, however, since the source is fixed far away from it, it is difficult to observe the obvious localisation on that hinge. More than that, the EM wave propagating to that corner also couple to the atmosphere outside the sample. Both of these two reasons prevent us from capturing obvious hinge localisation there. So in the text, only the hinge state localised on the hard boundary is shown.

In summary, we have designed and demonstrated a HODSM using the photonic crystal. The design is based on coupling 2D deformed honeycomb lattices with proper interlayer coupling along $z$ direction. The Dirac points emerge at the phase transition point between a HOTI  phase and a NI phase when $k_z$ varies. The hinge states that connect two Dirac points projected on a hinge are directly observed with field scanning measurement at microwave frequencies. Our work provides a concrete example of HODSM in photonic systems. Moreover, the design scheme presented here can also be applied to construct other types of photonic topological semimetals with higher-order band topology, such as Weyl semimetals and nodal line semimetals. 

We thank Y. Long for discussions. This work is supported by National Research Foundation Singapore Competitive Research Program No. NRF-CRP23-2019-0007.

\nocite{*}
\bibliographystyle{apsrev4-2}

\providecommand{\noopsort}[1]{}\providecommand{\singleletter}[1]{#1}%

\end{document}